\documentclass[10pt]{iopart}
\usepackage{threeparttable}
\usepackage{graphicx}
\begin{document}

\title[ ]{High-Density Superconductive Logic Circuits Utilizing 0 and $\pi$ Josephson Junctions}

\author{S Razmkhah and M Pedram}

\address{Department of Electrical and Computer Engineering, University of Southern California, Los Angeles, CA 90007 USA}
\ead{razmkhah@usc.edu}
\vspace{10pt}
\begin{indented}
\item[]August 2023
\end{indented}

\begin{abstract}
Superconductor Electronics (SCE) is a fast and power-efficient technology with great potential for overcoming conventional CMOS electronics' scaling limits.  
Nevertheless, the primary challenge confronting SCE today pertains to its integration level, which lags several orders of magnitude behind CMOS circuits. In this study, we have innovated and simulated a novel logic family grounded in the principles of phase shifts occurring in 0 and $\pi$ Josephson junctions.
The fast phase logic (FPL) eliminates the need for large inductor loops and shunt resistances by combining the half-flux and phase logic. Therefore, the Josephson junction (JJ) area only limits the integration density. The cells designed with this paradigm are fast, and the clock-to-Q delay is about 4ps while maintaining over 50\% parameter margins. This logic is power efficient and can increase the integration by at least 100$\times$ in the SCE chips. 
\end{abstract}

\vspace{2pc}
\noindent{\it Keywords}: Superconductor Electronics, VLSI, Single Flux Quantum, All-JJ\\
%\submitto{\SUST}
\maketitle
\ioptwocol

\section{Introduction}
flux quantum (SFQ) technology~\cite{Likharev1991} holds great promise for the next generation of very-large-scale integration (VLSI) circuits. Among SFQ logic circuits, rapid single flux quantum (RSFQ) stands out for its focus on high operation rates. RSFQ uses Josephson Junctions (JJs), which switch incredibly fast in just a few picoseconds (ps). RSFQ logic cells respond in about $10 \mathrm{ps}$ from clock to output, enabling RSFQ systems to work well at speeds between $40 \mathrm{GHz}$ and $60 \mathrm{GHz}$~\cite{RSFQ_Kato,RSFQ_Nagaoka}. The energy needed for a Josephson Junction to switch is much lower than in CMOS technology, even as low as $10^{-19}\, \mathrm{J/bit}$. This makes RSFQ systems more power-efficient than current technologies. Many studies focus on RSFQ technology, covering circuit and system designs~\cite{RSFQ_block_Hironaka,RSFQ_system_Kawaguchi,RSFQ_block_Cong}, layout designs~\cite{RSFQ_layout_Fourie,RSFQ_layout_Herbst}, as well as electronic design automation (EDA) tools and algorithms~\cite{RSFQ_EDA_Yang,RSFQ_EDA_Zhang2020}.

Despite the numerous merits of RSFQ circuits, they are not free of substantial challenges. Notably, the integration density of RSFQ circuits remains relatively modest, with approximately 10,000 logic gates accommodating a chip area of $1, \mathrm{cm^2}$. This scale of integration is inadequate to meet the computational requirements posed by today's demanding applications. RSFQ circuits face other limitations: the lack of compact on-chip memory solutions and the need for a substantial bias current for effective operation. These impediments underscore the significance of investigating alternative circuit families capable of surmounting these hurdles and propelling further advancements in SFQ technology.

The switching element in superconductor circuits is the Josephson Junction (JJ). JJ is an SFQ circuit's active component with a common Superconductor-Insulator-Superconductor (SIS) structure. The behavior of a JJ may be expressed by the Current-Phase Relationship (CPR):
\begin{equation} \label{eq:jj}
J_s(\phi) = J_{c}sin(\phi)
\end{equation}
where the $J_s$ is the current density of the JJ, $J_{c}$ is the critical current density of the JJ above which the JJ exits the superconducting state, and $\phi$ is the phase difference between two superconducting layers. This simplified CPR equation, which assumes that supercurrent always tunnels in the JJ's barrier and the temperature is below the critical temperature, approximates the JJ behavior well for  Superconductor-Insulator-Superconductor (SIS) JJs and is used in most SPICE-based simulator engines. 

The MITLL SFQ5ee process \cite{MITLL_Tolpygo2015,MITLL_Tolpygo2016} is one of the example technologies implementing a $\mathrm{Nb/Al-AlO_x/Nb}$ type junction where the material of superconducting layers is $Nb$ and the insulator is $AlO_{x}$. By replacing the barrier insulator layer with a magnetic material with a built-in magnetic field, the SIS JJ  becomes a magnetic junction (MJJ) \cite{MJJ_Ryazanov,MJJ_Baek}. The magnetic junction (MJJ) has been studied extensively for its unique characteristics, and some new devices based on the MJJ structure have been proposed, for example, the $\pi$-junction ($\pi$-JJ), $\phi$-junction ($\phi$-JJ) and $2\phi$-junction ($2\phi$-JJ). $\pi$-JJ \cite{MJJ_Ryazanov,piJJ_Guichard} has an intrinsic phase shift of $\pi$ as shown in Eq.\ref{eq:pi_jj}, which some researches have utilized to realize current saving designs \cite{piJJ_design_Ortlepp,piJJ_design_Yuki}.
\begin{equation} \label{eq:pi_jj}
J_s(\phi) = J_{c1}sin(\phi+\pi)
\end{equation}
Similarly, if the phase shift is not $\pi$ but an arbitrary value $\phi_0$, it forms the $\phi$-junction (Eq.\ref{eq:phi_jj}). The related works may be found in \cite{phi_JJ_Goldobin,phi_JJ_Pugach}.  
\begin{equation} \label{eq:phi_jj}
J_s(\phi) = J_{c1}sin(\phi+\phi_0)
\end{equation}

Increasing the on-chip density of SCE circuits is essential to their wider applicability. However, the growing mutual inductance and cross talk pose limitations on minimizing the dimensions of metal lines, although the kinetic inductors for passive transmission line (PTL) design may offer a potential solution. In an attempt to overcome the density challenge, Soloviev et al. \cite{2phi_design_Soloviev} developed logic cells (including NDRO, DRO, and half adder) utilizing $2\phi$-junctions, aiming to eliminate the need for inductors, and thus, enhance scalability. 
%Concurrently, efforts are underway to leverage the $2\phi$-junction to reduce the dynamic power consumption within an SFQ system. 
In their study, Salameh et al. \cite{2phi_design_Salameh} introduce three cells employing $2\phi$-junctions: a Josephson transmission line (JTL), an inverter, and an OR gate. Compared to conventional RSFQ cells, these cells employ half flux quantum (HFQ) pulses, reducing latency and switching power. Additionally, Hasegawa et al. \cite{RSFQ_half_Hasegawa} demonstrated an SFQ/HFQ interface circuit by combining 0- and $\pi$-Josephson junctions, although this implementation did not encompass the entirety of phase logic functionalities.

To shrink the circuit sizes even further, we design circuits using JJs with higher $J_C$ while eliminating the JJs' shunt resistances. High $J_C$ with self-shunted NbN/TaN/NbN JJs have been demonstrated in \cite{yan2022intrinsically}. SFS $\pi$ JJs have been demonstrated with NbN/PdNi/NbN. The JJs have $\sim$ 12 nm PdNi thickness, $J_C$ $\simeq$ 1000 $\mathrm{\mu A/\mu m^2}$ \cite{pham2022weak}. Others have proposed phase logic based on $\pi$ JJs \cite{maksimovskaya2022phase}.

This paper introduces a standard cell library that leverages $\pi$-junctions to implement FPL cells, aiming to reduce the footprint of superconductive logic. Logic cells are showcased, demonstrating a remarkable size reduction of at least 100$\times$ compared to standard SFQ cells. The PTLs are reduced in width due to an increase in the circuit impedance to $\mathrm{\sim 1.8\mu m}$. The diverse range of cells provided caters to the fundamental needs of various computing systems. The functionality of these cells is verified using the JoSIM \cite{JoSIM_Delport} simulator, while optimization using the qCS tool \cite{AltayQCS2022} yields satisfactory margins. The paper presents critical circuit parameters for FPL cells. Note that the projected layout area for cells is computed assuming that the SIS $J_C$ is 600 $\mathrm{\mu A/\mu m^2}$ and that for the SFS $J_C$ is 1000 $\mathrm{\mu A/\mu m^2}$.
\section{All-JJ Circuits}
\subsection{$2\phi$-Junction}
\label{sec:2phi_jj}
\noindent
Recently, there have been works showing that at the $0-\pi$ transition, the fundamental term of the CPR vanishes, making the high-order harmonic terms non-negligible \cite{phi_JJ_Goldobin}. Moreover, in \cite{2phi_Stoutimore}, a single SFS junction using the $Cu_{47}Ni_{53}$ alloy barrier was implemented with two parallel superconducting inductors: a readout inductor and a small shunt inductor. The readout inductor is coupled to a commercial DC superconducting quantum interference device (SQUID) sensor, which detects flux $\Phi$ in the readout loop. By measuring the CPR on different barrier thicknesses in different temperatures, reference \cite{2phi_Stoutimore} demonstrated a $\pi$-periodic behavior, eliminating any other alternative explanations but having a second-order CPR. Thus, the overall CPR may be rewritten to be:
\begin{equation} \label{eq:2phi}
J_s(\phi) = J_{c1}sin(\phi)+J_{c2}sin(2\phi)
\end{equation}
And a new device named $2\phi$-JJ comes to light with the CPR shown as:
\begin{equation} \label{eq:2phi_jj}
J_s(\phi) = J_{c2}sin(2\phi)
\end{equation}

Several intriguing observations can be made regarding the $2\phi$-JJ ($2\phi$-JJ). Firstly, its CPR follows a period of $\pi$ instead of the more typical period of $2\pi$. Secondly, the $2\phi$-JJ undergoes switching when a $\pi$ phase jump occurs, generating a half flux quantum ($\frac{1}{2}\Phi_0=1.03\times10^{-15}Wb$). Consequently, each switching event of a $2\phi$-JJ corresponds to a phase shift of $\pi$. A few studies have been reported utilizing $2\phi$ junctions, as detailed below.

\subsection{Replacing $2\phi$ with 0 and $\pi$ JJs}
\begin{figure}[ht]
    \includegraphics[width=0.2\textwidth]{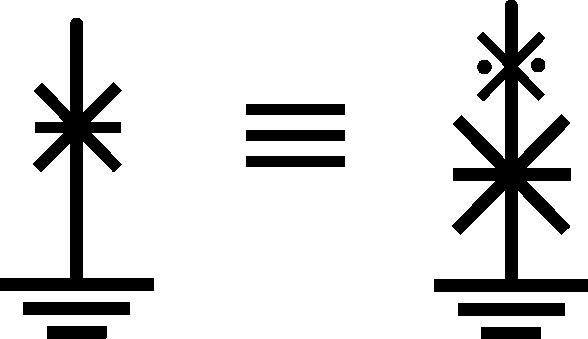}
    \centering
    \caption{Replacing a switching $\pi$ JJ with a switching 0 JJ and a series non-switching $\pi$ JJ that only provides phase shift.}
    \centering
    \label{fig:switch}
\end{figure}

\noindent
Some logic cells designed with $2\phi$ JJs were presented in \cite{2phi_design_Salameh}. Unfortunately, a dependable fabrication process for these junctions remains lacking. An innovative new approach presented by Soloviev et al. \cite{soloviev2022pi} demonstrated the feasibility of implementing $2\phi$ JJs using only 0 and $\pi$ JJs. This procedure combines 0 and $\pi$ JJs to create a bistable structure, functioning analogously to a $2\phi$ JJ. However, a potential concern arises from using both 0 and $\pi$ JJs as switching elements within this structure, possibly impacting the circuit reliability. In case of switching happens in both SIS and SFS layers, different parameter variations between two layers can cause unreliable switching and reduce the margins. Therefore, in this work, we design the cells such that all the switching happens in only $\pi$ JJs.
 In case a design needs switching in both layers, refinement can be made by substituting the switching $\pi$ JJ with a 0 JJ in series with a higher $I_{C\pi} > 2\times I_{C0}$ $\pi$ JJ, as illustrated in fig.~\ref{fig:switch}. This modification ensures that all switching actions occur within the SIS (0 JJ) layer, thus enhancing overall reliability.

\section{Fast Phase Logic (FPL)}
\noindent

This work introduces a collection of novel superconductive logic cells designed to admit compact layouts. This logic family leverages high critical current density ($J_C$) 0 and $\pi$ Josephson Junctions (JJs) to establish an ALL-JJ-based superconductive logic cell family, named \textit{fast phase logic} (FPL). Notably, the JJs within the FPL family operate without the need for shunt resistance. Moreover, no explicit inductances are present in the design of logic cells. This results in very high layout density.

For example, the proposed JTL cell within this paradigm occupies a mere 0.8 $\mathrm{\mu m^2}$ in size and incorporates four JJs, with only the $\pi$ JJ serving as a switching element. The power consumed during switching these JJs is approximately $\mathrm{3\times 10^{-20} W}$. This indicates that a chip spanning $1 \mathrm{cm^2}$ and housing approximately $\mathrm{2.5 \times 10^7}$ JJs would consume around 97 mW in the worst-case scenario, making it amenable for cooling using liquid helium.

\begin{figure}[ht]
    \includegraphics[width=0.45\textwidth]{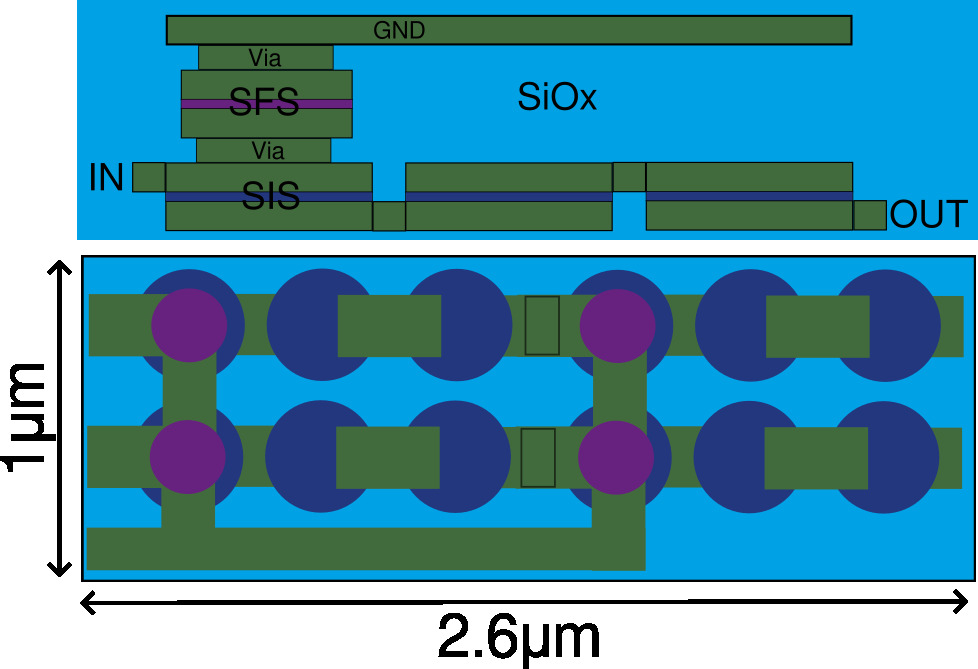}
    \centering
    \caption{A layout sample of the FPL cells assuming the high $J_C$ SFS and SIS technology. Here we assume that SIS JJs have $J_C = 600 \mu A/\mu m^2$ and SFS JJs have $J_C = 1000 \mu A/\mu m^2$}
    \centering
    \label{fig:FPL}
\end{figure}

A visual representation of the layout featuring four JTL cells employing the FPL paradigm can be observed in Fig.~\ref{fig:FPL}. Evident from the figure, the four JTL cells presented here, alongside the bias line, would occupy 2.6 $\mathrm{\mu m^2}$ of the chip area. Compared to the RSFQ library, where each JTL cell occupies a minimum of $\mathrm{20 \times 20\mu m^2}$, the FPL cells are approximately $\sim 500\times$ smaller.

\subsection{Logic Cell Implementation with FPL}
\label{sec:standard_cell}
\noindent
In this design approach, most cells follow the similar conventional architecture used in SFQ, with storing loops that determine the cell state and non-storing loops that propagate the signal. The $\pi$ JJs are switching elements. 0 JJs propagate the signal as inductances while a combination of 0 and $\pi$ JJs store is used for storing half flux. As a result, the logic '1' within this logic family is depicted through a half flux quantum pulse, where the product of voltage and time corresponds to half of a full flux quantum ($\frac{1}{2}\Phi_0=1.03\times10^{-15}Wb$).

Even though this design removes the need for explicit inductors, minor parasitic inductors are inherent in the circuit due to the interconnections and vias between SIS, SFS, and ground (GND) layers. A conservative assumption has been made throughout the design process, attributing a $0.1pH$ inductance to each connection within the same layer and a $0.3pH$ inductance to vias. Notably, these inductances are omitted from the circuit schematics for simplicity. Following margin calculations, the outcomes indicate that these parasitic elements will exert negligible influence on the circuits until they exceed a value of $>$1pH.

Numerous logic cells were designed, and their corresponding circuits were simulated using the JoSIM software. The simulations incorporated thermal noise, although its impact remained negligible due to the absence of shunts in the JJs. Remarkably, these circuits exhibit enhanced resistance to flux trapping, as the absence of inductive loops prevents flux coupling with the circuits.
This paper showcases a selection of exemplary cells tailored for elevated margins. These basic cells illustrate the efficient implementation of fast and dependable logic through the FPL approach.

\subsection{Wiring cells}
\subsubsection{Josephson transmission line (JTL)}
\begin{figure}[ht]
    \includegraphics[width=0.3\textwidth]{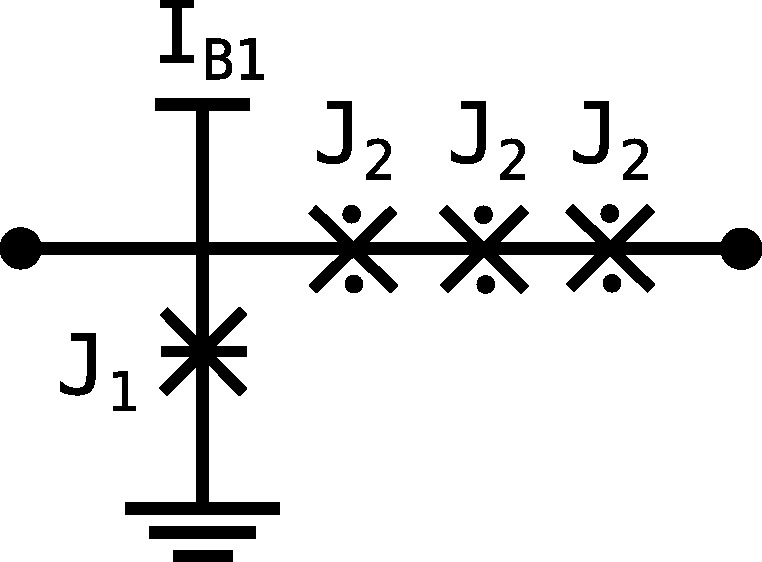}
    \centering
    \caption{Schematic of the Josephson Transmission Line (JTL). Here $J_1$ is 30$\mu A$, $J_2$ is 43$\mu A$, and $I_{B1}$ is 15$\mu A$. After optimization, the values change to 22.3$\mu A$,36.1$\mu A$, and 13.6$\mu A$, respectively.}
    \centering
    \label{fig:JTL_sch}
\end{figure}

\noindent
Fig.\ref{fig:JTL_sch} illustrates the schematic of a JTL cell. Within this schematic, $J_1$ denotes the switching JJ, while three junctions labeled $J_2$ function as inductance components for the JTL cell. The JTL cell is a foundational component in the design, responsible for interconnecting various other blocks and ensuring impedance matching.

JTLs exhibit cascading capabilities, meaning that linking the OUT port of one JJ with the IN port of a successive JTL establishes a two-element JTL chain. The collaboration between J1-J2 and the subsequent JTL constructs a closed loop, establishing a phase equation in which integrating phase differences across the loop equates to an integer multiple of $2\pi$. When a half-flux-quantum (HFQ) pulse enters the IN port, J1 switches, engendering another HFQ pulse that propagates to the subsequent device. This sequence facilitates the transport of HFQ pulses along the JTL.

\begin{figure}[ht]
    \includegraphics[width=0.4\textwidth]{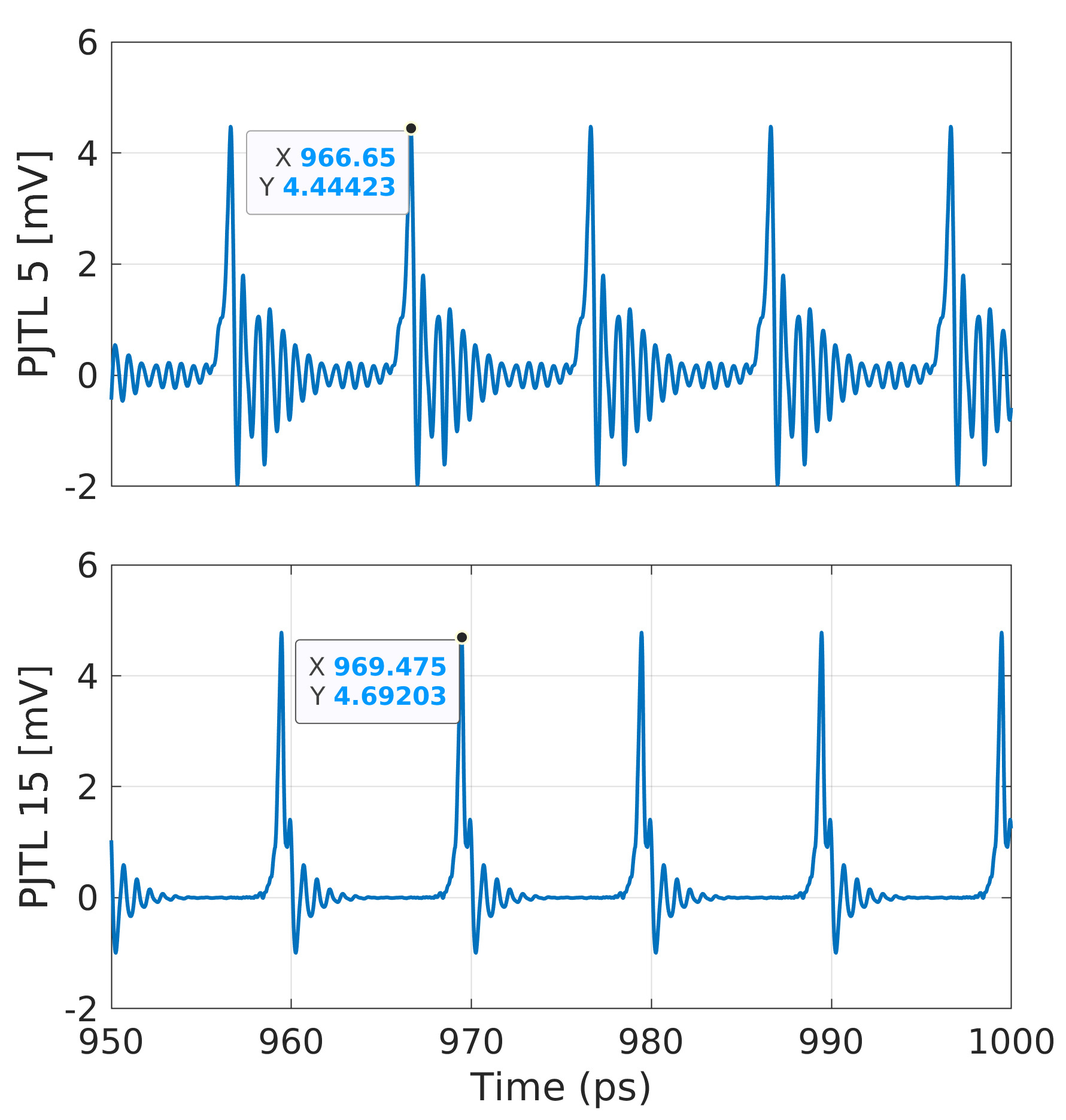}
    \centering
    \caption{Simulation waveform of 15 series JTLs.}
    \centering
    \label{fig:JTL_sim}
\end{figure}

The waveform captured through simulations is portrayed in Fig.\ref{fig:JTL_sim}. Notably, the simulated waveform for a chain of 15 interconnected JTLs indicates that each JTL cell introduces a delay of approximately 0.3 ps. The output of this setup interfaces with a load that converts the HFQ pulses into SFQ pulses.

\begin{figure}[ht]
    \includegraphics[width=0.4\textwidth]{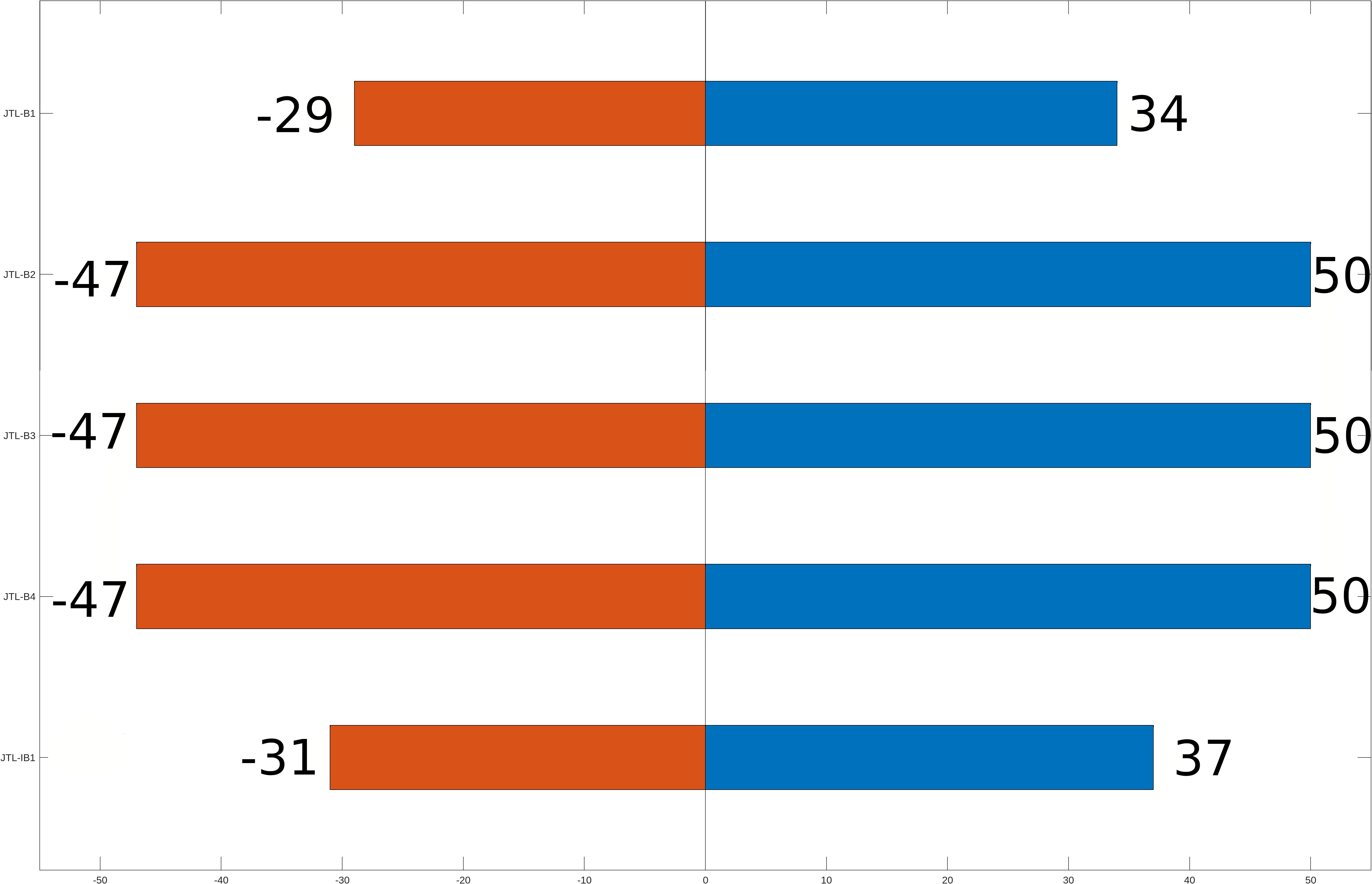}
    \centering
    \caption{Parameter margin of the JTL cell after optimization. After a few iteration cycles, about 60\% margin was achieved.}
    \centering
    \label{fig:JTL_optim}
\end{figure}

\subsection{DC/FPL converter}
\begin{figure}[ht]
    \includegraphics[width=0.45\textwidth]{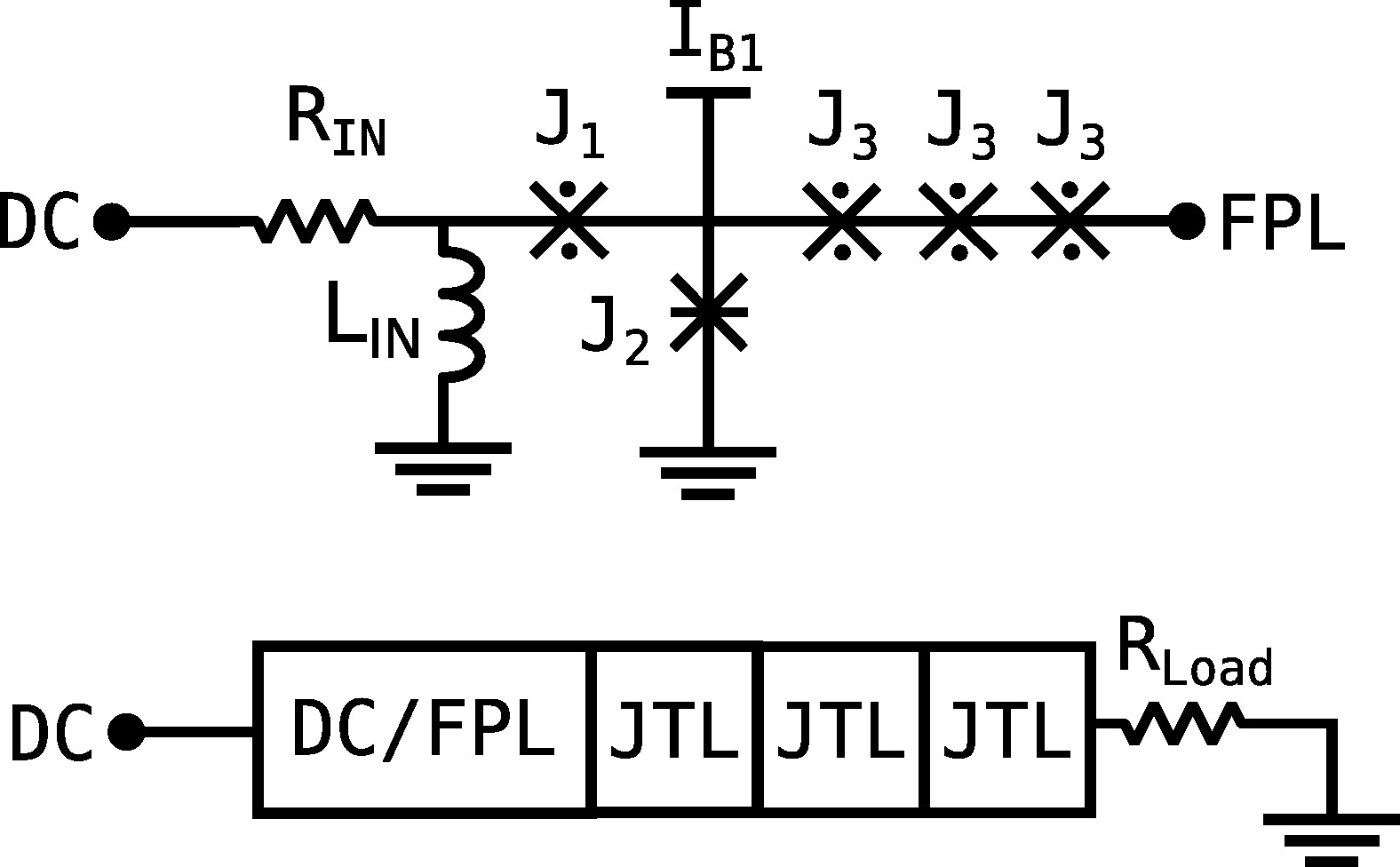}
    \centering
    \caption{Schematic of the DC/SFQ converter.  Here $R_{IN} = 50 \Omega$, $L_{IN} = 1pH$, $J_1 = 48\mu A$, $J_2 = 30\mu A$, $J_3 = 42\mu A$ and $I_{B1}$ is 10$\mu A$.}
    \centering
    \label{fig:DCSFQ_sch}
\end{figure}

\begin{figure}[ht]
    \includegraphics[width=0.4\textwidth]{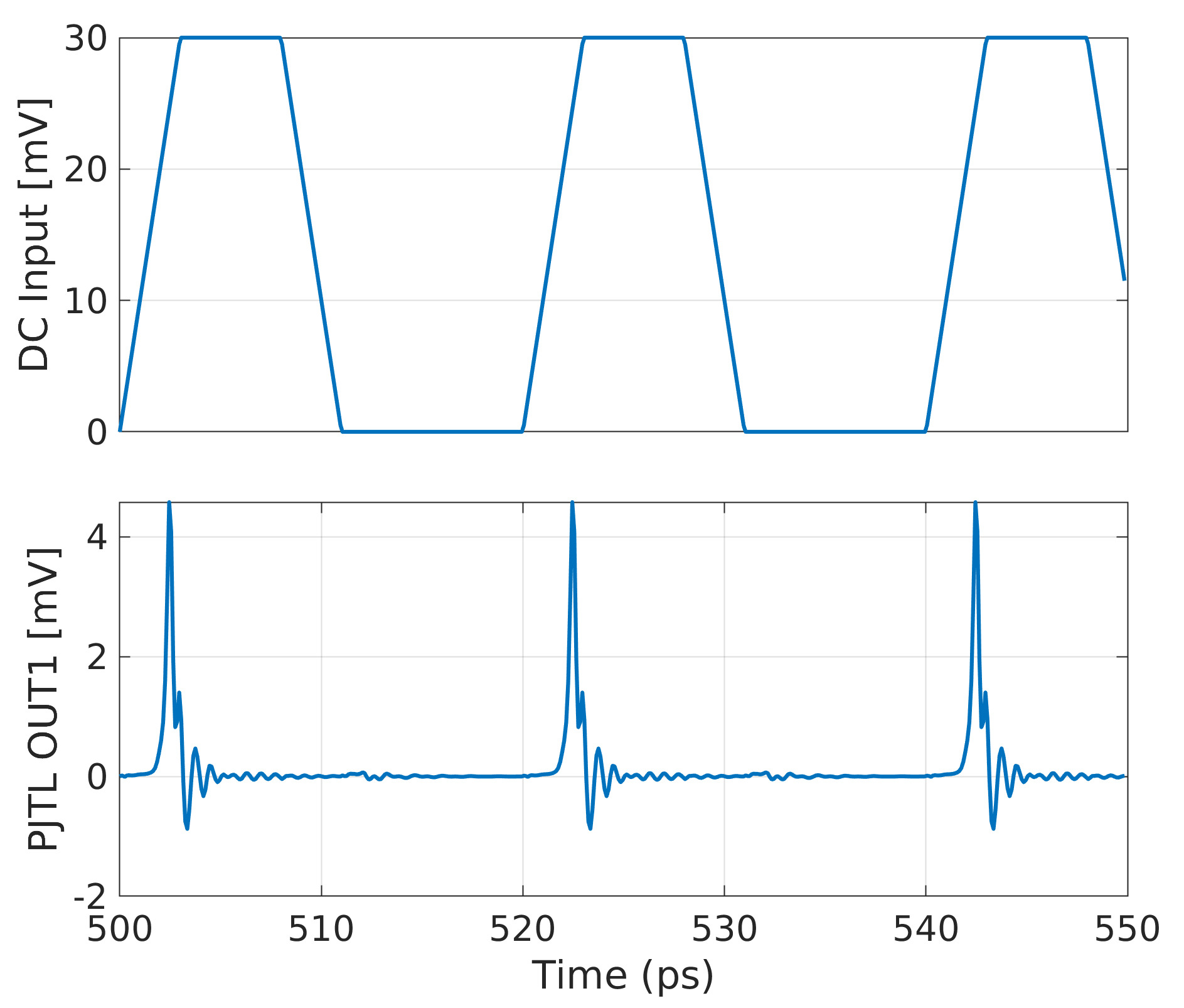}
    \centering
    \caption{Simulation waveform of the DC/SFQ converter.}
    \centering
    \label{fig:DCSFQ_sim}
\end{figure}

\noindent
Fig.\ref{fig:DCSFQ_sch} presents the schematic and component values of a DC/FPL converter. Within this configuration, $R_{IN}$ signifies a serial 50 $\Omega$ input resistor designed to transform the input voltage into a current. This resistor can be implemented either on-chip (as per this design) or off-chip. An inductor $L_{IN}$ is introduced following the input resistor. During the rising edge of the input signal, $L_{IN}$ exhibits high impedance, leading the majority of current to flow through J2. As a consequence, an FPL pulse is triggered at the output port.
Upon stabilization of the input voltage, $L_{IN}$ operates as a short connection, diverting the input current through $L_1$ towards the ground. This flow spares J2 from activation. The simulated waveform is depicted in Fig.\ref{fig:DCSFQ_sim}, elucidating the dynamic behavior of the converter.

\subsection{SPLITTER}
\begin{figure}[ht]
    \includegraphics[width=0.45\textwidth]{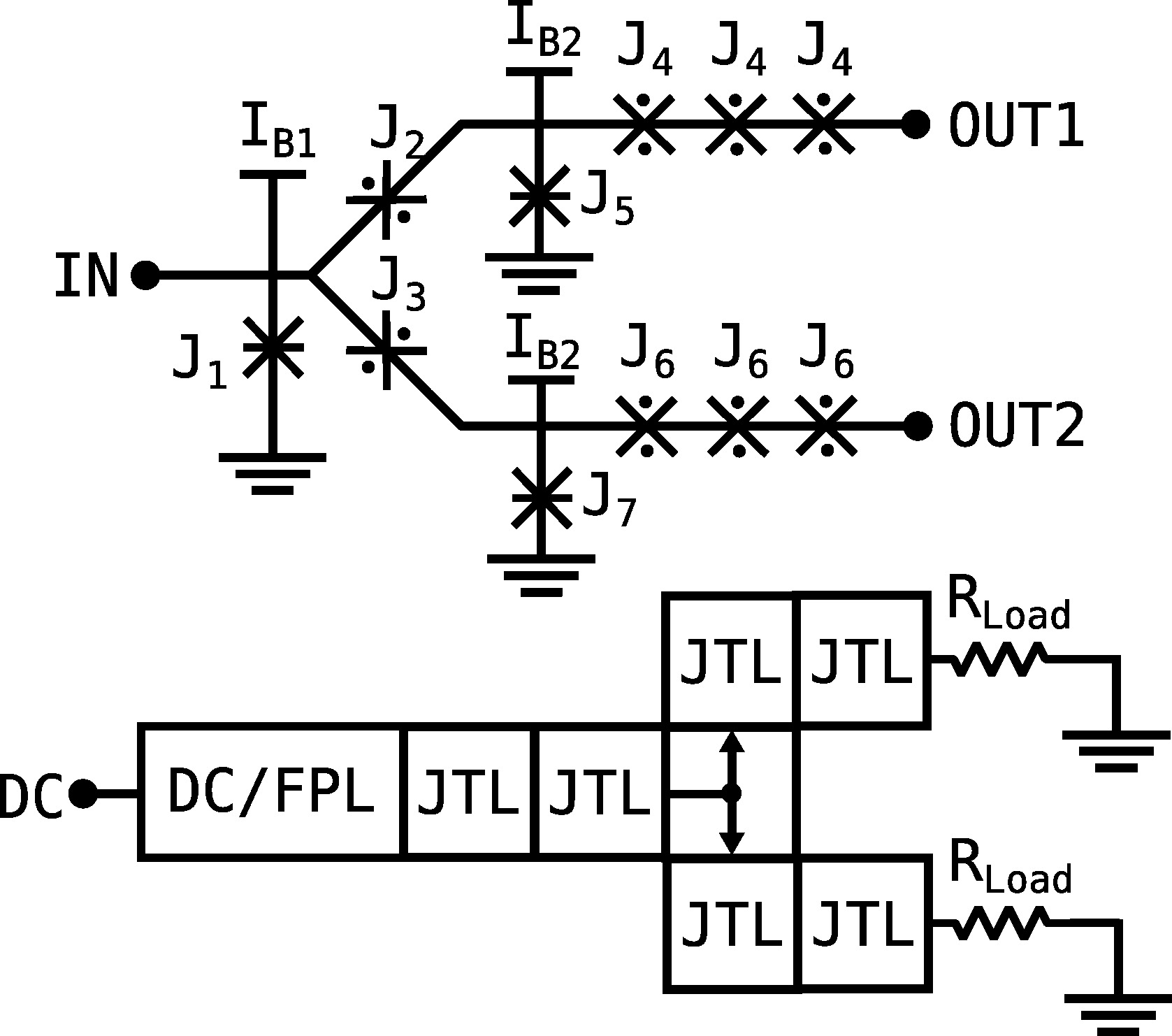}
    \centering
    \caption{Schematic of the splitter cell. Here $J_1 = 30\mu A$, $J_2 = 24\mu A$, $J_3 = 24\mu A$, $J_4 = 39\mu A$, $J_5 = 27\mu A$, $J_6 = 39\mu A$, $J_7 = 27\mu A$ and $I_{B1} = 25\mu A, I_{B2} = 15\mu A$}
    \centering
    \label{fig:Splitter_sch}
\end{figure}

\noindent
Like SFQ logic cells, the FPL cells also exhibit a fan-out of one. To address this limitation, a splitter is employed to replicate the pulse. Fig.\ref{fig:Splitter_sch} illustrates the schematic of the splitter cell, featuring its corresponding component values. The associated test circuit is also depicted, highlighting its architecture. In this configuration, $J_1$ receives the pulse through the IN port. The looping current undergoes division, effectively triggering J5 and J4 independently. This outcome leads to distinct FPL pulses at each output port.

\begin{figure}[ht]
    \includegraphics[width=0.4\textwidth]{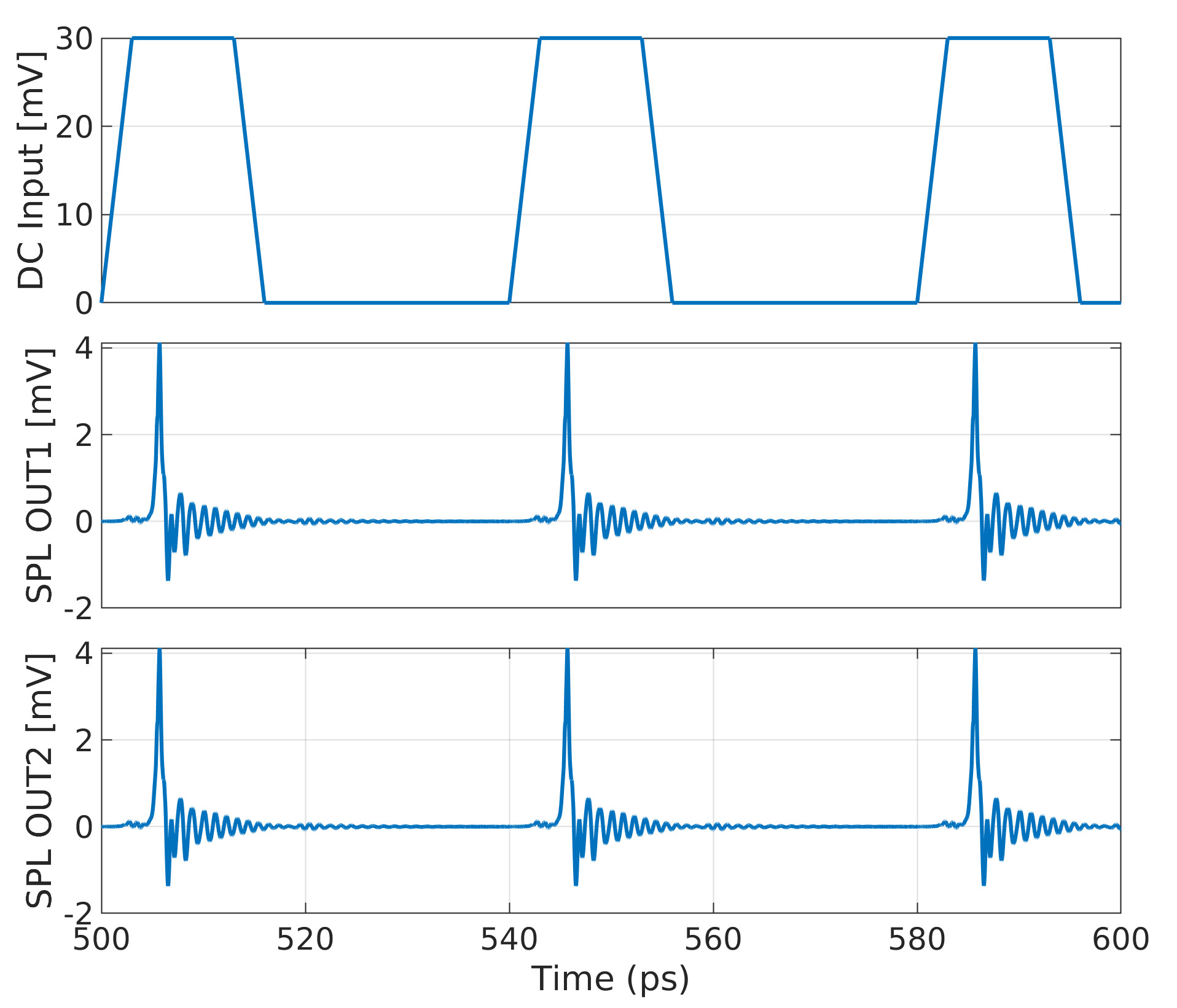}
    \centering
    \caption{Simulation waveform of the splitter cell.}
    \centering
    \label{fig:splitter_sim}
\end{figure}

The simulation waveform in Fig.\ref{fig:splitter_sim} showcases the dynamic behavior. Notably, the input signal derives from the DC to DC/FPL cell, and subsequent passage through the JTL cells results in a bifurcation of the pulse. The readout pulses on the respective loads are also portrayed within this illustration.

\subsection{Merger (asynchronous OR)}
\begin{figure}[ht]
    \includegraphics[width=0.45\textwidth]{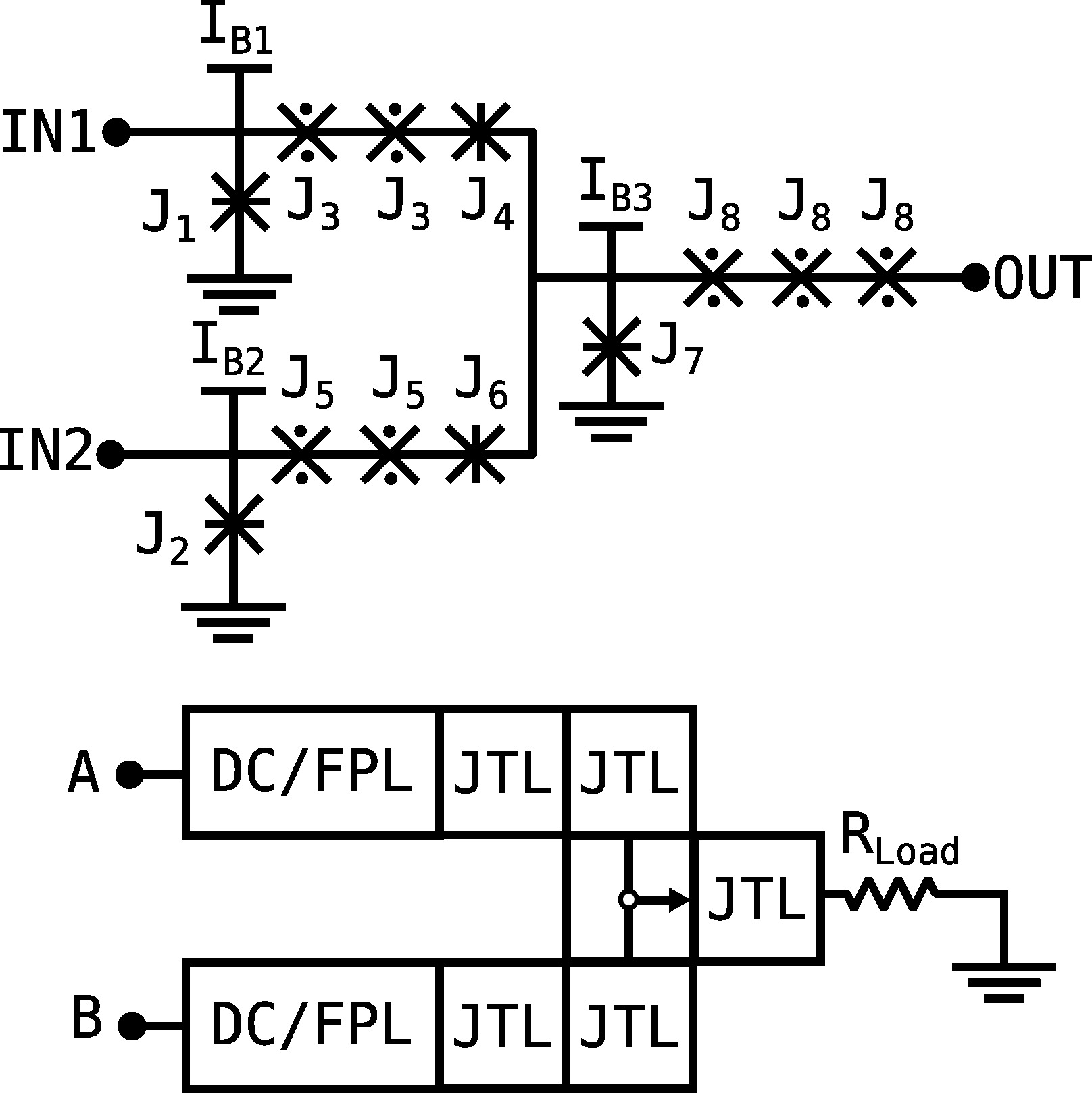}
    \centering
    \caption{Schematic and test circuit of the merger cell. Here $J_1 = 33\mu A$, $J_2 = 33\mu A$, $J_3 = 66\mu A$, $J_4 = 25\mu A$, $J_5 = 66\mu A$, $J_6 = 25\mu A$, $J_7 = 25\mu A$, $J_8 = 48\mu A$ and $I_{B1} = 15\mu A, I_{B2} = 15\mu A, I_{B3} = 27\mu A$}
    \centering
    \label{fig:Merger_sch}
\end{figure}

\noindent
Fig.\ref{fig:Merger_sch} shows the schematic of the merger cell, alternatively referred to as the confluence buffer or asynchronous OR gate. Upon reception of an FPL pulse from either input port (e.g., $IN_1$), the FPL triggers the receiving junction ($J_1$). This action culminates in amplifying the current within the corresponding branch ($J_1$-$_3$-$J_4$-$J_7$). Consequently, $J_7$ switches, generating an output pulse.
Simultaneously, on the alternate branch ($J_6$ in the case of input from IN1), the buffering junction is activated to neutralize the backward-flowing flux toward the other input port (IN2). Specifically, $J_4$ and $J_6$ collectively prevent the propagation of pulses in the reverse direction. 

\begin{figure}[ht]
    \includegraphics[width=0.4\textwidth]{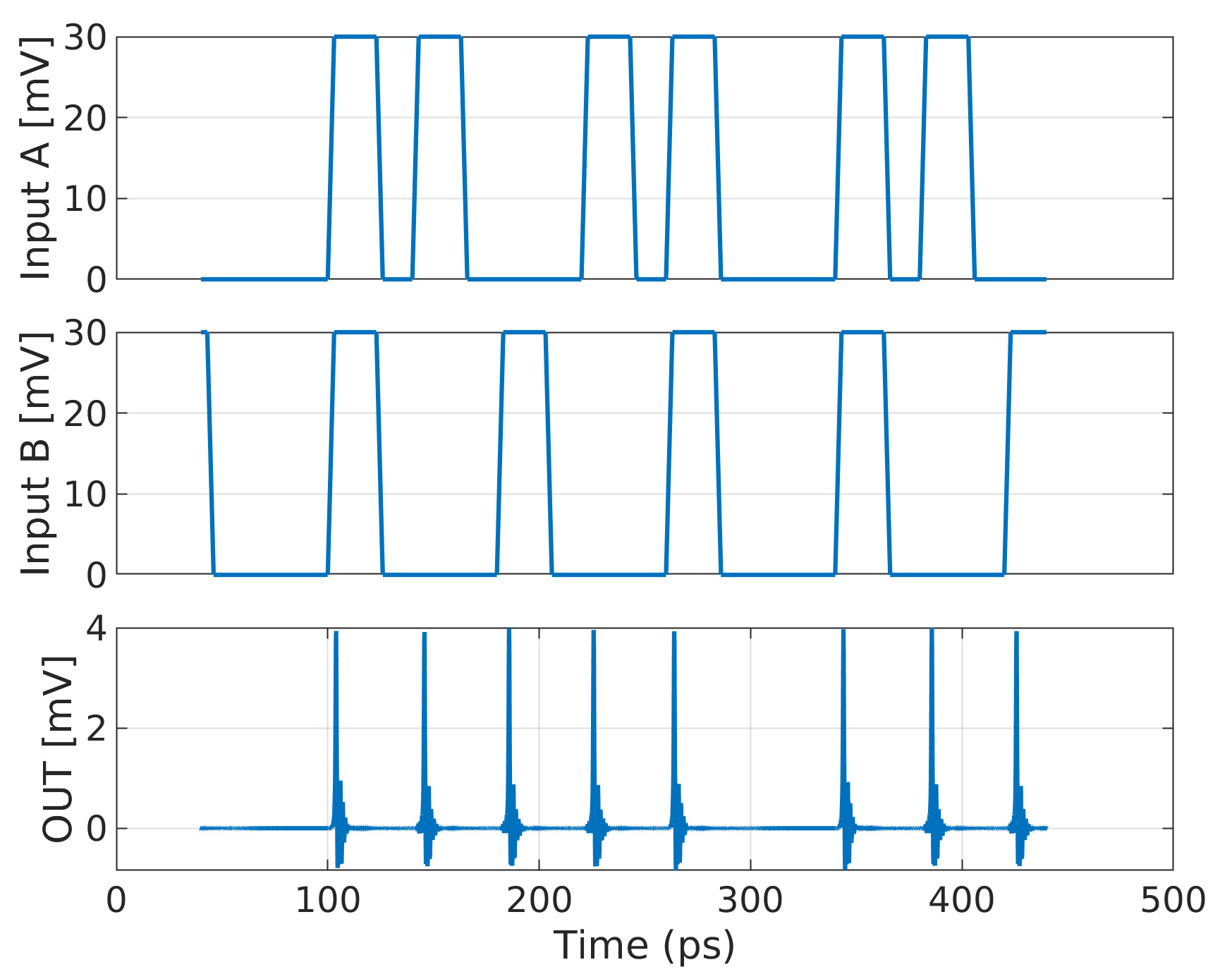}
    \centering
    \caption{Simulation waveform of the merger cell.}
    \centering
    \label{fig:Merger_sim}
\end{figure}

In instances where two input pulses coincide or fall within a narrow temporal window (a few picoseconds), only one output pulse is emitted at the OUT port. Fig.\ref{fig:Merger_sim} captures the simulation waveform, illustrating the input and output waveforms and the intricate dynamics of this configuration.

\subsection{D flip-flop}
\begin{figure}[ht]
    \includegraphics[width=0.45\textwidth]{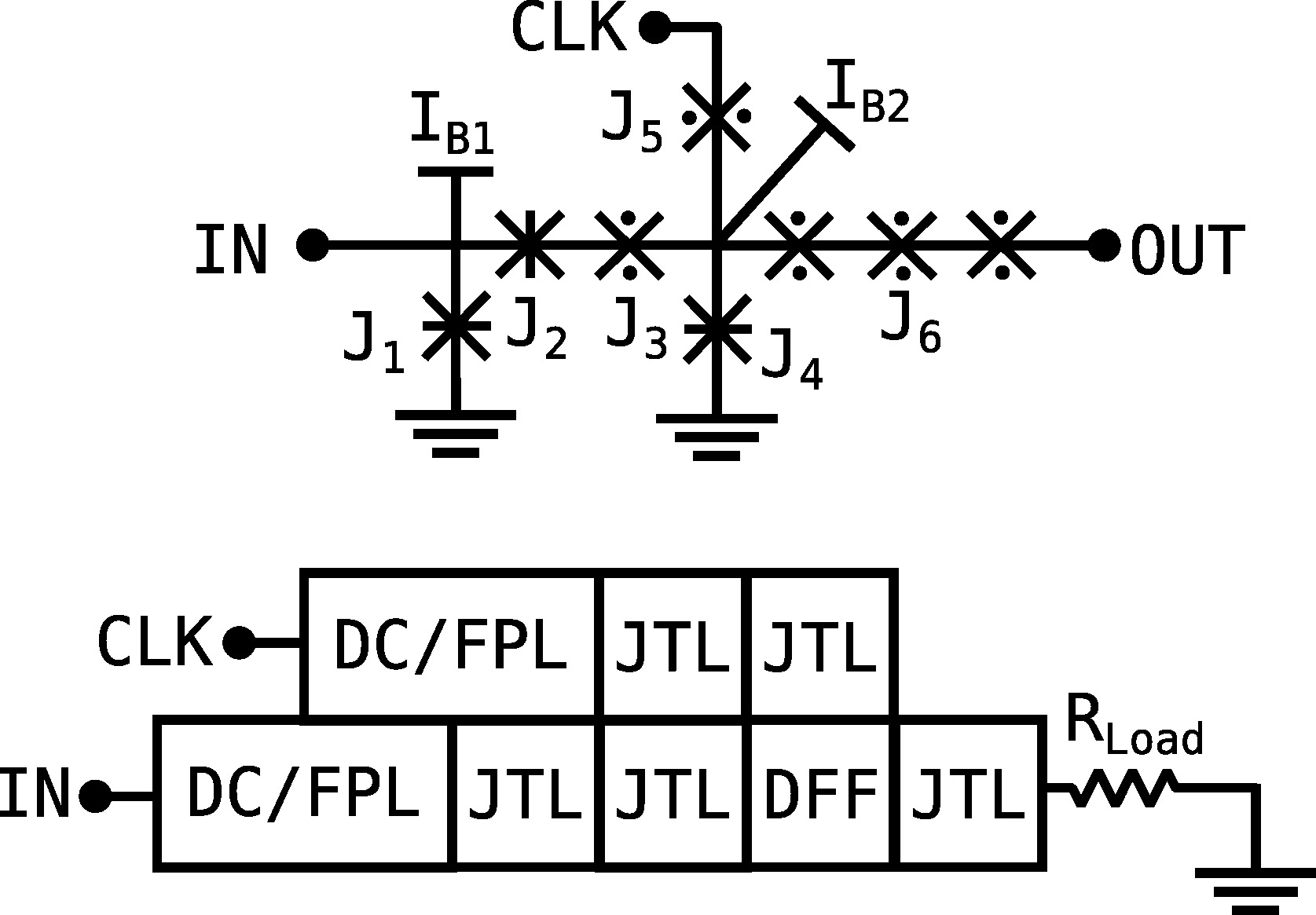}
    \centering
    \caption{Schematic of the D Flip-flop. Here $J_1 = 30\mu A$, $J_2 = 33\mu A$, $J_3 = 66\mu A$, $J_4 = 30\mu A$, $J_5 = 32\mu A$, $J_6 = 54\mu A$, and $I_{B1} = 10\mu A, I_{B2} = 18\mu A$}
    \centering
    \label{fig:DFF_sch}
\end{figure}

\noindent
Fig.\ref{fig:DFF_sch} introduces the schematic and corresponding test bench of the DFF (D Flip-Flop) cell. When an FPL pulse arrives through the IN port, the pulse is preserved within the $J_1$-$_2$-$J_3$-$J_4$ loop as a clockwise screening current. This action concurrently heightens the bias current of $J_4$. Consequently, when the incoming FPL pulse from the CLK port materializes, $J_4$ is triggered, engendering an FPL pulse at the output port.

\begin{figure}[ht]
    \includegraphics[width=0.4\textwidth]{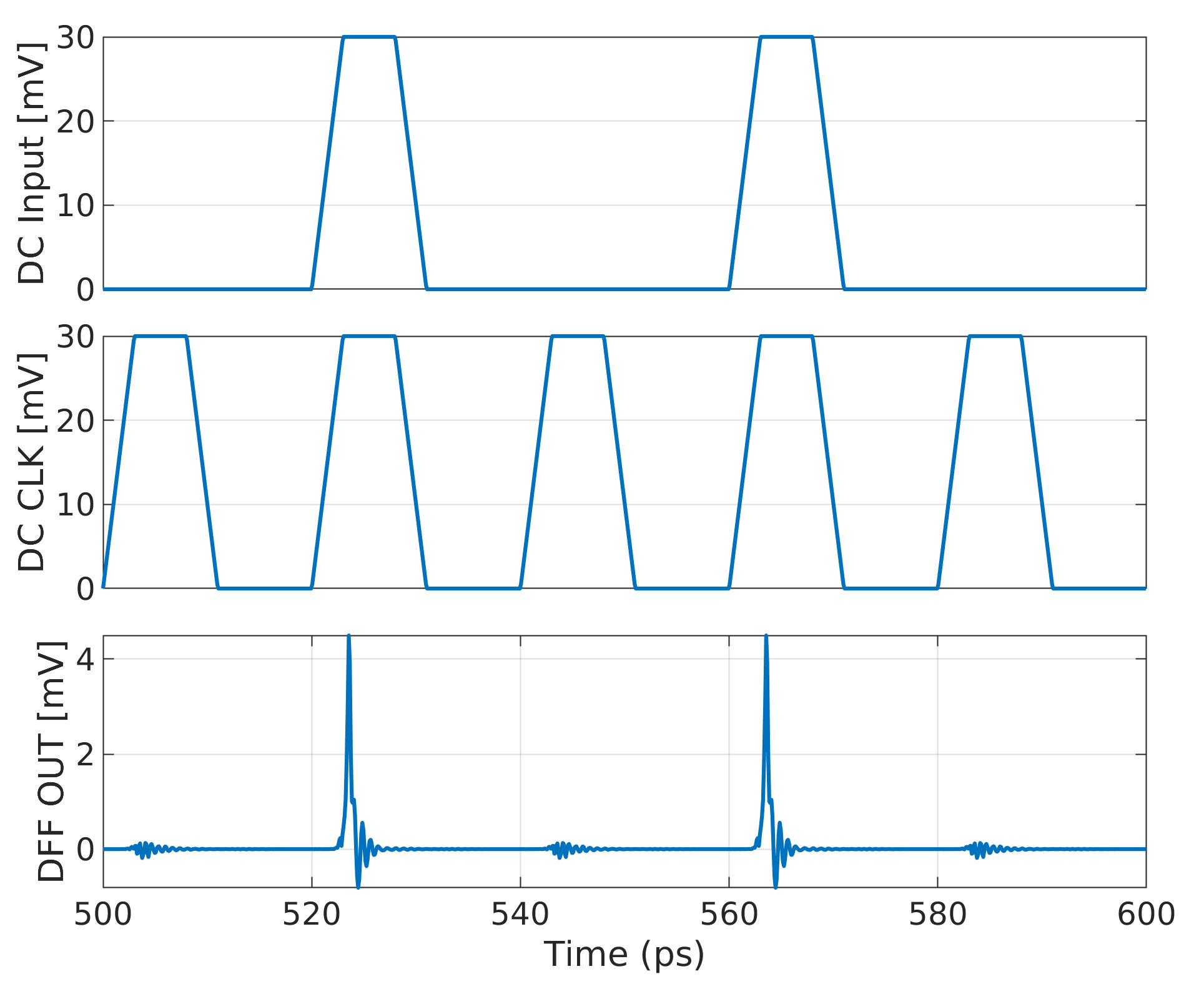}
    \centering
    \caption{Simulation waveform of the D Flip-flop.}
    \centering
    \label{fig:DFF_sim}
\end{figure}

Contrastingly, in the absence of a stored pulse, the pulse incoming from CLK activates $J_5$, while $J_4$ remains untouched. The pulse energy is then dissipated into the ground. The simulation waveform, showcased in Fig.\ref{fig:DFF_sim}, affords a comprehensive depiction of this process, illustrating the input and output waveforms and the underlying mechanisms at play.

\subsection{Asynchronus AND gate}
\begin{figure}[ht]
    \includegraphics[width=0.45\textwidth]{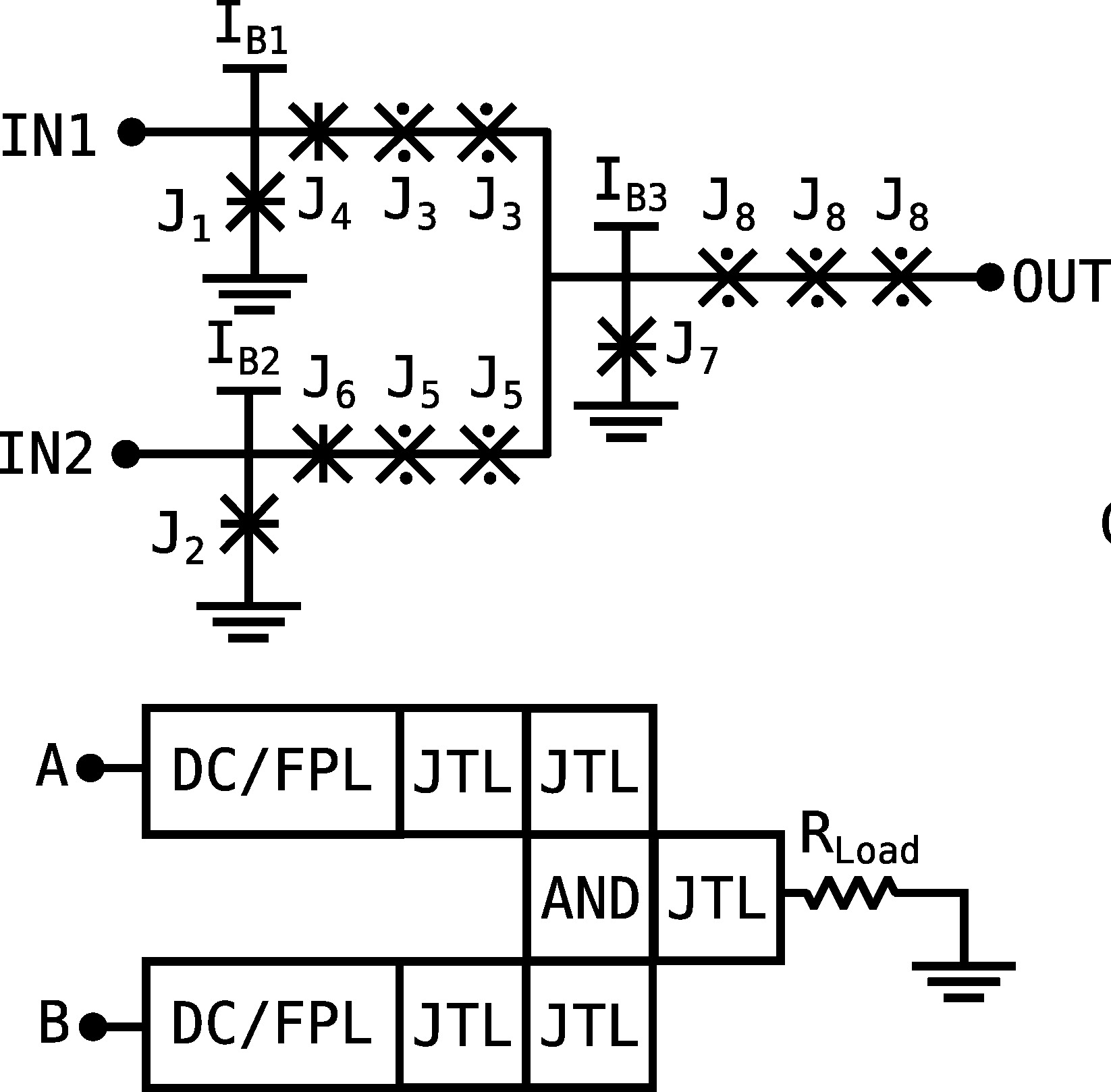}
    \centering
    \caption{Schematic of the AND gate. Here $J_1 = 35\mu A$, $J_2 = 35\mu A$, $J_3 = 72\mu A$, $J_4 = 25\mu A$, $J_5 = 72\mu A$, $J_6 = 25\mu A$, and $I_{B1} = 15\mu A, I_{B2} = 15\mu A, I_{B3} = 10\mu A$}
    \centering
    \label{fig:AND_sch}
\end{figure}

\begin{figure}[ht]
    \includegraphics[width=0.4\textwidth]{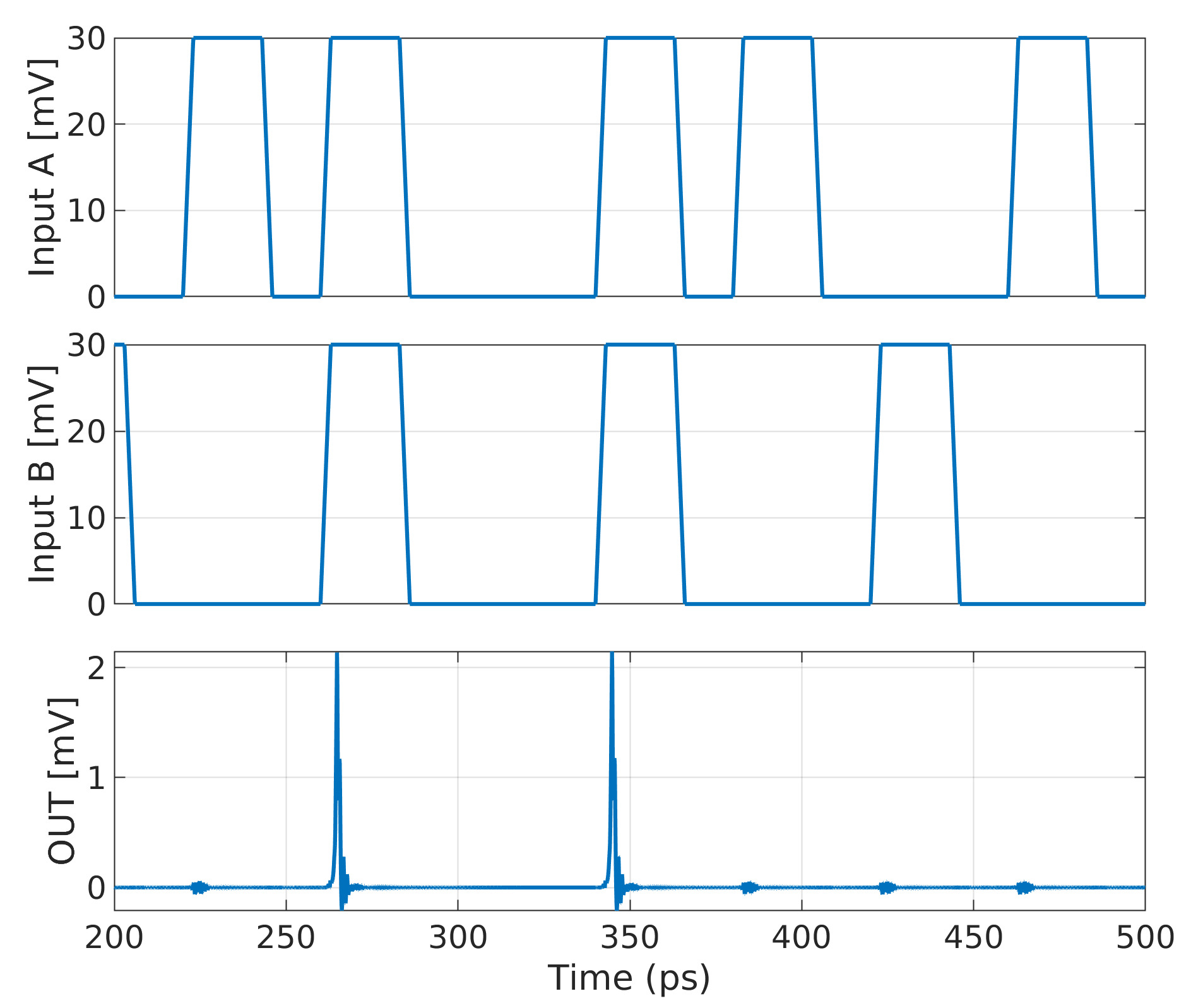}
    \centering
    \caption{Simulation waveform of the AND gate.}
    \centering
    \label{fig:AND_sim}
\end{figure}

\noindent
The AND gate employs a structure akin to an OR (Merger) gate, with adjustments made to the key components (merging part). Consequently, the output junction $J_7$ requires a minimum of two FPL pulses to generate an output signal. Fig.\ref{fig:AND_sch} presents the AND gate configuration with its associated test circuit and the listed components as indicated in the caption. The simulation waveform is displayed in Fig.\ref{fig:AND_sim}, portraying the process, including both the input-output waveforms and the underlying mechanisms.

\subsection{Clocked AND/OR gates}
\begin{figure}[ht]
    \includegraphics[width=0.35\textwidth]{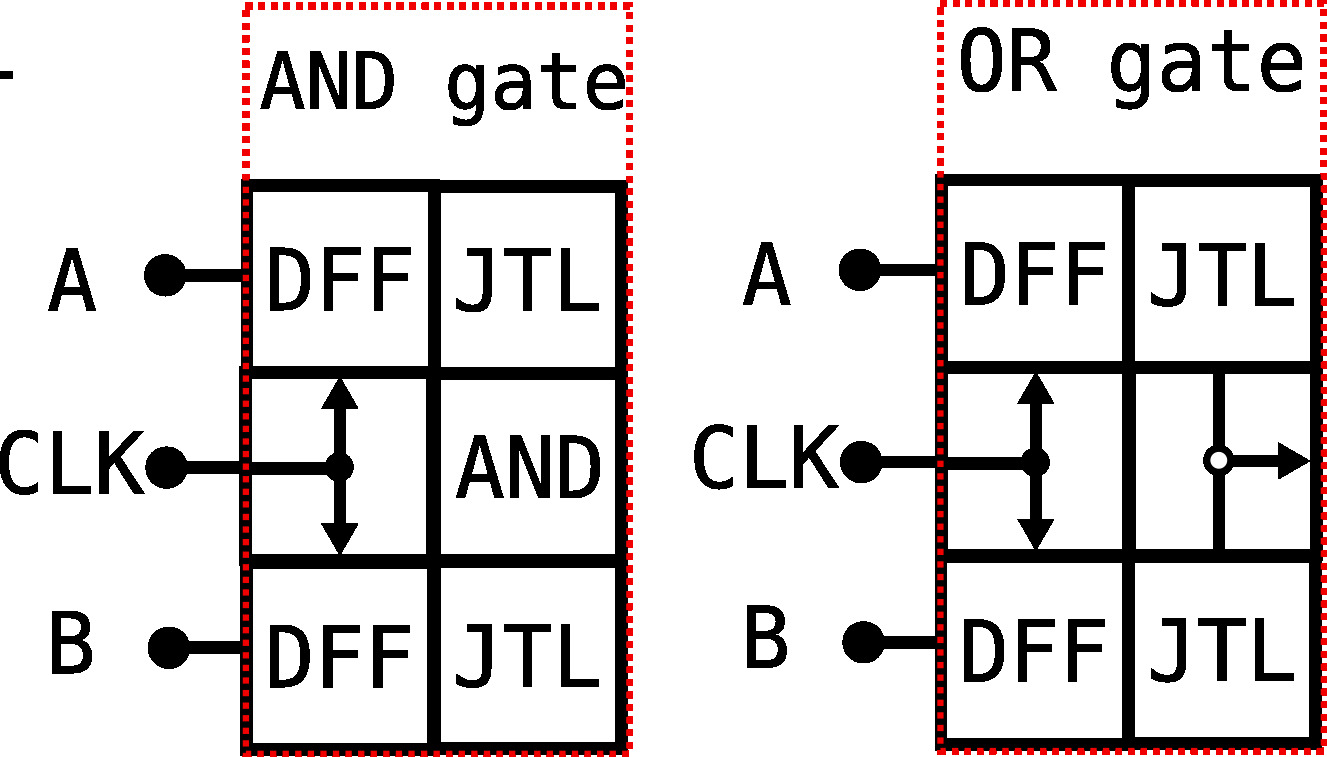}
    \centering
    \caption{Schematic of the OR gate.}
    \centering
    \label{fig:CLK_sch}
\end{figure}

\noindent
As is well-known, clocked gates can be derived from the asynchronous gates by incorporating DFFs at their inputs. For instance, the OR gate is constructed by integrating two DFFs at each input of a merger cell. These DFFs receive synchronized clock signals from a splitter circuit. The same fundamental configuration can be adapted to transform the asynchronous AND gate into a clock-controlled variant. The schematics of these two types of gates are depicted in Fig.\ref{fig:CLK_sch}.

These structures are amenable to optimization, allowing for the enhancement of margins, reduction in cell sizes, and minimization of Q-to-Clock propagation times. Such optimizations can lead to improved overall performance and efficiency in practical applications.

These presented cells serve as illustrative examples, showcasing the potential of FPL technology to facilitate the creation of compact, high-speed, and reliable logic circuits. The relevant details are summarized in Table~\ref{table:delays}, which provides insights into the designed cell delays, the maximum simulated frequency, and the projected size.

In estimating the size, we consider the JJ technology with demonstrated values of $J_C = 600 \mu A/\mu m^2$ for the SIS layer and $J_C = 1000 \mu A/\mu m^2$ for the SFS layer. This estimation suggests that FPL cells can be integrated at densities of up to 50 $\mathrm{MJJ/cm^2}$ and can operate efficiently at clock frequencies of up to 50 GHz. This demonstrates the potential for FPL technology to offer scalability and high-performance capabilities.

\begin{table}
\caption{\label{table:delays}Delay and estimated size of each designed cell from input-output/ clock-to-Q}
\footnotesize\rm
\begin{tabular}{@{}llll}
\br
Cell & Delay (ps) & Test freq. (GHz) & Cell size ($\mathrm{\mu m^2}$) \\
\mr
     JTL & 0.28 & 125 & 0.65\\
     SPL & 0.35 & 100 & 1.8\\
     Merger & 1 & 90 & 2\\
     DFF & 0.75 & 90 & 2.1\\
     AND & 2.7 & 90 & 9.3\\
     OR & 2.95 & 90 & 9.3\\
     DC/FPL$^*$ & 1.1 & 100 & 1.4\\
\br
\end{tabular}
\begin{tablenotes}
      \small
      \item $^*$ The delay is calculated from threshold to pulse and depends on the input value. Size is without input resistor.
\end{tablenotes}
\end{table}

\section{Conclusion}
\label{sec:conc}
\noindent
The FPL logic cells were demonstrated, harnessing the power of 0 and $\pi$-JJs. Operating without the need for inductors (akin to $2\phi$ logic) and unburdened by shunts (like HFQ logic), these cells represent a groundbreaking development. The emergence of this novel logic family opens up avenues for unprecedented miniaturization within SCE logic. Leveraging an inventive design, these cells demand bias currents that are a mere $20\times$ % lower than conventional RSFQ cells, accompanied by an astonishing size reduction of at least $100\times$ %.

Remarkably, the absence of inductive loops renders FPL logic cells considerably less susceptible to trapped fluxes and crosstalk. This sets the FPL logic family apart from other SFQ technologies. Capitalizing on the potential for dependable dense integration and high operational frequencies, the FPL logic family emerges as a promising contender poised to shape the trajectory of the next generation of Very Large Scale Integration (VLSI) circuits.

\ack
\noindent
This work was partly supported by the National Science Foundation (NSF) through the project Expedition: DISCoVER (Design and Integration of Superconducting Computation for Ventures beyond Exascale Realization) under Grant 2124453.
The authors thank Holing Cong for designing a $2\phi$ cell library and Dr. D. S. Holmes for his valuable comments.

\end{document}